\documentclass[amsmath,amssymb,aps,eqsecnum,twocolumn,tightenlines,floats,floatfix,prc,nofootinbib,preprintnumbers]{revtex4}

\def\st#1{{\kern-4pt} \not\!#1}

\def\sp{\kern +3pt}
\def\sm{\kern -3pt}

\usepackage[usenames]{color}

\newcommand{\GeV}{\mathrm{GeV}}
\newcommand{\dd}{\mathrm{d}}
\newcommand{\ii}{\mathrm{i}}

\usepackage{graphics}
\usepackage{graphicx}
\usepackage{epsf}
\usepackage{amsmath}
\usepackage{amssymb}
\usepackage{slashed}
\usepackage{bm}

\begin{document}

\phantom{0}
\vspace{-0.2in}
\hspace{5.5in}

\preprint{}

\vspace{-1in}

\title
{\bf
R\^ole of the pion electromagnetic form factor
in the $\Delta(1232) \to \gamma^\ast N$  \\
timelike transition}
\author{G.~Ramalho$^{1}$, M.~T.~Pe\~na$^{2}$,
J.~Weil$^{3}$,  H.~van Hees$^{3,4}$
and U.~Mosel$^5$
\vspace{-0.1in}  }

\affiliation{
$^1$International Institute of Physics, Federal
University of Rio Grande do Norte, Avenida Odilon Gomes de Lima 1722,
Capim Macio, Natal-RN 59078-400, Brazil
\vspace{-0.15in}}
\affiliation{
$^2$Centro de F\'{i}sica Te\'orica e de Part\'{i}culas (CFTP),
Instituto Superior T\'ecnico (IST),
Universidade de Lisboa,
Avenida Rovisco Pais, 1049-001 Lisboa, Portugal
\vspace{-0.15in}}
\affiliation{
$^3$Frankfurt Institute for Advanced Studies, Ruth-Moufang-Stra{\ss}e 1,
D-60438 Frankfurt, Germany
\vspace{-0.15in}}
\affiliation{
$^4$Institut f{\"u}r Theoretische Physik, Universit{\"a}t Frankfurt,
Max-von-Laue-Stra{\ss}e 1, D-60438 Frankfurt, Germany
\vspace{-0.15in}}
\affiliation{
$^5$Institut f{\"u}r Theoretische Physik, Universitaet Giessen,
D35392- Giessen, Germany}

\vspace{0.2in}
\date{\today}

\phantom{0}

\begin{abstract}
  The $\Delta(1232)  \to \gamma^\ast N$ magnetic dipole form factor
  ($G_M^\ast$) is described here within a new covariant model that
  combines the valence quark core together with the pion cloud
  contributions.
  The pion cloud term is parameterized by two terms: one connected to
  the pion electromagnetic form factor, the other to the photon
  interaction with intermediate baryon states.  The model can be used in
  studies of pp and heavy ion collisions.  In the timelike region this
  new model improves the results obtained with a constant form factor
  model fixed at its value at zero momentum transfer.  At the same time,
  and in contrast to the Iachello model, this new model predicts a peak
  for the transition form factor at the expected position, i.e. at the
  $\rho$ mass pole.
  We calculate the decay of the $\Delta \to \gamma N$ transition, the
  Dalitz decay ($\Delta \to e^+ e^- N$), and the $\Delta$ mass
  distribution function.
  The impact of the model on dilepton spectra in pp collisions is also
  discussed.
\end{abstract}

\vspace*{0.9in}  
\maketitle

\section{Introduction}

To understand the structure of hadrons, baryons in particular, in terms
of quarks and gluons at low energies, is theoretically challenging due
to the intricate combination of confinement and spontaneous chiral
symmetry breaking, and the non-perturbative character of QCD in that
energy regime. Fortunately, experimentally electromagnetic and hadron
beams in accelerator facilities are decisive tools to reveal that
structure, and seem to indicate a picture where effective degrees of
freedom as baryon quark cores dressed by clouds of mesons play an
important role. For a review on these issues see~\cite{NSTAR}.  Although
different, experiments with electromagnetic and strong probes complement
each other. In electron scattering, virtual photons disclose the region
of momentum transfer $q^2<0$, and spacelike form factors are obtained
\cite{NSTAR,MesonBeams,HADES14}. Scattering experiments of pions or
nucleons with nucleon targets involving Dalitz decays of baryon
resonances \cite{MesonBeams,Timelike,HADES14,Frohlich10} provide
information on timelike form factors, defined in the $q^2>0$ region
where the meson spectrum arises.
The results of all these different measurements
have to match at the photon point ($q^2=0$).

Among the several baryon resonances the $\Delta$ excitation and decays
have a special role and are not yet fully understood.  The
electromagnetic transition between the nucleon and the $\Delta(1232)$,
and in particular its dominant magnetic dipole form factor
$G_M^\ast(q^2)$, as function of $q^2$, is a prime example that discloses
the complexity of the electromagnetic structure of the excited states of
the nucleon and illustrates the limitations of taking into account only
valence quark degrees of freedom for the description of the transition.

In the region of small momentum transfer $G_M^\ast(q^2)$ is usually
underestimated by valence quark contributions alone.  Several models
have been proposed in order to interpret this finding.  Most of them are
based on the interplay between valence quark degrees of freedom and the
so-called meson cloud effects, in particular, the dominant pion cloud
contribution~\cite{NSTAR,NDelta,NDeltaD,LatticeD,Octet2Decuplet,Octet2Decuplet2,DynamicalModels}.
Other recent works on the $\Delta \to \gamma^\ast N$ transition can be
found in Refs.~\cite{Braun06,Alexandrou11,Eichmann12,Segovia14}.

In this work we propose a hybrid model 
which combines the valence quark component, determined by a constituent
quark model, constrained by lattice QCD and indirectly by experimental
data, with a pion cloud component. The pion cloud component is written
in terms of the pion electromagnetic form factor and therefore
constrained by data.

The  $\Delta \to \gamma^\ast N$ transition in the timelike region was
studied using vector meson dominance (VMD)
models~\cite{Schafer94b,deJong96,Krivoruchenko02,Faessler03}, the constant form factor
model~\cite{Zetenyi03,Frohlich10}, a two component model (model with
valence quark and meson cloud decomposition), hereafter called the
Iachello model~\cite{Frohlich10,Iachello,Iachello73}, and the covariant
spectator quark model~\cite{Timelike} (which incidentally also assumes
VMD for the quark electromagnetic current).

The Iachello model pioneered the timelike region studies of the
 $\Delta \to \gamma^\ast N$ transition. The model was successful in the
description of the nucleon form factors~\cite{Iachello73} but has been
criticized for generating the pole associated with the $\rho$-meson pole
near $q^2 \simeq 0.3$ GeV$^2$, below
$q^2= m_\rho^2 \simeq 0.6 \; \GeV^2$~\cite{Frohlich10} as it should. The
constant form factor model is a good starting point very close to
$q^2=0$ but, on the other hand, does not satisfactorily take into account
the finite size of the baryons and their structure of non-pointlike
particles.

In the covariant spectator quark model the contributions for the
transition form factors can be separated into valence quark and meson
cloud effects (dominated by the pion). The valence quark component is
directly constrained by lattice QCD data, and has been seen to coincide
with the valence quark core contributions obtained from an extensive
data analysis of pion photo-production~\cite{Lattice,LatticeD,EBAC}.
Its comparison to experimental data enables the extraction of
information on the complementary meson cloud component in the spacelike
region~\cite{Timelike,NDelta}.  However the extension to the timelike
region of the meson cloud is problematic given the difficulty of a
calculation that comprises also in a consistent way the whole meson
spectrum. In Ref.~\cite{Timelike} the meson cloud was parameterized by a
function $F_\rho$, taken from the Iachello model where it describes the
dressing of the $\rho$-propagator by intermediate $\pi \pi$ states.
As noted before, unfortunately, the function $F_\rho$ has a peak that is
displaced relatively to the $\rho$-meson pole mass.
Here, by directly using the pion form factor data we corrected for this
deficiency.

\begin{figure}[t]
\vspace{.4cm}
\includegraphics[width=2.6in]{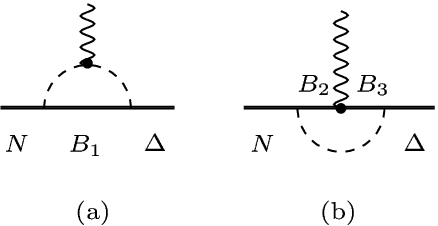}
\caption{Pion cloud contributions for the $\Delta \to \gamma^\ast N$
  electromagnetic transition form factors.  Between the initial and
  final state there are several possible intermediate octet baryon
  and/or decuplet baryon states: $B_1$ in diagram (a); $B_2$ and $B_3$
  in diagram (b).}
\label{figMC}
\end{figure}

Moreover, in previous
works~\cite{Timelike,NDelta,NDeltaD,LatticeD,Octet2Decuplet} we have
assumed that the pion cloud contributions for the magnetic dipole form
factor could be represented by a simple parameterization of one term
only. But in the present work we introduce an alternative
parameterization of the pion cloud which contains two terms. These two
leading order contributions for the pion cloud correspond to the two
diagrams of Fig.~\ref{figMC}.  We use then a parameterization of the
pion cloud contributions for $G_M^\ast$ where diagram (a) is related to
the pion electromagnetic form factor $F_\pi (q^2)$, and is separated
from diagram (b). Diagram (a), where the photon couples directly to the
pion, is dominant according to chiral perturbation theory, which is
valid in the limit of massless and structureless quarks. But the other
contribution, from diagram (b), where the photon couples to intermediate
(octet or decuplet) baryon states
while the pion is exchanged  between those states,
becomes relevant in models with constituent quarks with dressed masses
and non-zero anomalous magnetic moments.
This was shown in Ref.~\cite{Octet2Decuplet2} on the study of the meson
cloud contributions to the magnetic dipole moments of the octet to
decuplet transitions. The results obtained for the
$\Delta \to \gamma^\ast N$ transition in particular, suggests that both
diagrams contribute with almost an equal weight.

\section{Iachello model}

In the Iachello model the dominant contribution to the
$\Delta \to \gamma^\ast N$ magnetic dipole form factor is the meson
cloud component (99.7\%) \cite{Frohlich10}. The meson cloud
contributions is estimated by VMD in terms of a function $F_\rho$ from
the dressed $\rho$ propagator, which in the limit $q^2 \gg 4 m_\pi^2$,
reads~\cite{Timelike}
\begin{equation}
\begin{split}
  F_{\rho}(q^2) &= \frac{m_\rho^2}{m_\rho^2 - q^2 - \frac{1}{\pi}
    \frac{\Gamma_\rho^0}{m_\pi}q^2 \log \frac{q^2}{m_\pi^2}
    + i \frac{\Gamma_\rho^0}{m_\pi} q^2}, \\
  &= \frac{m_\rho^2}{m_\rho^2 + Q^2 + \frac{1}{\pi}
    \frac{\Gamma_\rho^0}{m_\pi} Q^2 \log \frac{Q^2}{m_\pi^2}}.
\label{eqFrho}
\end{split}
\end{equation}
In the previous equation $Q^2=-q^2$, $m_\pi$ is the pion mass, and
$\Gamma_\rho^0$ is a parameter that can be fixed by the experimental
$\rho$ decay width into $2 \pi$, $\Gamma_\rho^0=0.149$ GeV or
$\Gamma_\rho^0=0.112$ GeV depending on the specific model
\cite{Timelike,Iachello}.

\section{Covariant spectator quark model}

Within the covariant spectator quark model framework the nucleon and the
$\Delta$ are dominated by the $S$-wave components of the quark-diquark
configuration~\cite{Nucleon,NDelta,Nucleon2}.  In this case the only
non-vanishing form factor of the $\Delta \to \gamma^\ast N$ transition
is the magnetic dipole form factor, which anyway dominates in all
circumstances.

One can then write~\cite{NDelta,NDeltaD,LatticeD}
\begin{equation}
G_M^\ast(q^2,W) = G_M^B(q^2,W) + G_M^\pi (q^2),
\end{equation}
where $G_M^B$ is the contribution from the bare core and $G_M^\pi$ the
contribution of the pion cloud. Here $W$ generalizes the $\Delta$ mass
$M_\Delta$ to an arbitrary invariant mass $W$ in the intermediate
states~\cite{Timelike}.  We omitted the argument $W$ on $G_M^\pi$ since
we take that function to be independent of $W$.

Following Refs.~\cite{NDelta,NDeltaD,LatticeD,Timelike}
we can write
\begin{equation}
G_M^B(q^2,W) =
\frac{8}{3 \sqrt{3}} \frac{M}{M+ W} f_v(q^2)
{\cal I}(q^2,W),
\label{eqGMbare}
\end{equation}
where 
\begin{equation}
{\cal I}(q^2,W)= \int_k \psi_\Delta (P_+,k) \psi_N (P_-,k),
\end{equation}
is the overlap integral of the nucleon and the $\Delta$ radial wave
functions which depend on the nucleon ($P_-$), the Delta ($P_+$) and the
intermediate diquark ($k$) momenta.  The integration symbol indicates
the covariant integration over the diquark on-shell momentum.  For
details see Refs.~\cite{NDelta,Timelike}.

As for $f_v(q^2)$ it is given by
\begin{equation}
f_v(q^2)= f_{1-}(q^2) + \frac{W + M}{2M} f_{2-}(q^2)
\end{equation}
where $f_{i-}$ ($i=1,2$) are the quark isovector form factors that
parameterize the electromagnetic photon-quark coupling. The form of this
parameterization assumes VMD mechanism~\cite{Nucleon,NDelta,Omega}. See
details in Appendix~\ref{appQuarkFF}.

In this work we write the pion cloud contribution as
\begin{equation}
G_M^\pi (q^2)=
3 \frac{\lambda_\pi}{2}
\left[
F_{\pi} (q^2) \left( \frac{\Lambda_\pi^2}{\Lambda_\pi^2 - q^2}
\right)^2 + \tilde G_D^2 (q^2)
\right],
\label{eqGMmc}
\end{equation}
where $\lambda_\pi$ is a parameter that define the strength of the pion
cloud contributions, $F_{\pi} (q^2)$ is a parameterization of the pion
electromagnetic form factor and $\Lambda_\pi$ is the cutoff of the pion
cloud component from diagram (a). The function $\tilde G_D^2$ on
Eq.~(\ref{eqGMmc}) simulates the contributions from the diagram (b), and
therefore includes the contributions from several intermediate
electromagnetic transitions between octet and/or decuplet baryon states.

From perturbative QCD arguments it is expected that the latter effects
fall off with $1/Q^8$~\cite{Carlson}. At high $Q^2$ a baryon-meson
system can be interpreted as a system with $N=5$ constituents, which
produces transition form factors dominated by the contributions of the
order $1/(Q^2)^{(N-1)}= 1/Q^8$. This falloff power law motivates our
choice for the form of $\tilde G_D^2$: the timelike generalization of a
dipole form factor
$G_D = \left(\frac{\Lambda_D^2}{\Lambda_D^2 - q^2} \right)^2$, where
$\Lambda_D$ is a cutoff parameter defining the mass scale of the
intermediate baryons.

The equal relative weight of the two terms of Eq.~(\ref{eqGMmc}), given
by the factor $\frac{1}{2}\lambda_\pi$, was motivated by the results
from Ref.~\cite{Octet2Decuplet2}, where it was shown that the
contribution from each diagram (a) and (b) for the total pion cloud in
the  $\Delta \to \gamma^\ast N$ transition is about 50\%.
The overall factor 3 was included for convenience, such that in the
limit $q^2=0$ one has $G_M^\pi (0)= 3 \lambda_\pi$. Since
$G_M^\ast(0) \simeq 3$, $\lambda_\pi$ represents the fraction of the
pion cloud contribution to $G_M^\ast(0)$.

In the spacelike regime, in order to describe the valence quark behavior
($1/Q^4$) of the form factors associated with the nucleon and $\Delta$
baryons, the dipole form factor $G_D$ with a cutoff squared value
$\Lambda_D^2 = 0.71 \, \GeV^2$
had been used
in previous works~\cite{Nucleon,NDelta}.
As we will show, a model with $\Lambda_D^2 = 0.71 \, \GeV^2$ provides a
very good description of the $\Delta \to \gamma^\ast N$
form factor data in the
region $-2 \, \GeV^2 < q^2 < 0$.
However, since in the present work we are focused on the timelike
region, we investigate the possibility of using a larger value for
$\Lambda_D^2$, such that the effects of heavier resonances
($\Lambda_D^2 \approx 1 \, \GeV^2$) can also be taken into account.

To generalize $G_D$ to the timelike region we define $\tilde G_D (q^2)$
\begin{equation}
\tilde G_D (q^2) =
\frac{\Lambda_D^4}{(\Lambda_D^2-q^2)^2  + \Lambda_D^2 \Gamma_D^2 },
\end{equation}
where $\Gamma_D (q^2)$ is an effective width discussed in
Appendix~\ref{appRegularize}, introduced to avoid the pole
$q^2= \Lambda_D^2$.  Since $\Gamma_D (0)=0$, in the limit $q^2=0$, we
recover the spacelike limit $\tilde G_D (0) = G_D(0)=1$. We note that
differently from the previous work \cite{Timelike} $\Tilde G_D$ is the
absolute value of $G_D$, and not its real and imaginary parts together.

To summarize this Section: Eq.~(\ref{eqGMmc}) modifies the expression of
the pion cloud contribution from our previous works, by including an
explicit term for diagram (b) of Fig.~\ref{figMC}. Diagram (a) is
calculated from the pion form factor experimental data.  Diagram (b)
concerns less known phenomenological input.  The $q^2$ dependence of
that component is modeled by a dipole function squared. 
Since $\lambda_\pi$ was fixed already by the low $q^2$ data, in the
spacelike region, the pion cloud contribution is defined only by the two
cutoff parameters $\Lambda_\pi$ and $\Lambda_D$.

Next we discuss the parameterization of the
pion electromagnetic form factor $F_\pi (q^2)$, which fixes the term
for diagram (a) and is
known experimentally.

\section{Parameterization of $F_\pi (q^2)$}

The data associated with the pion electromagnetic form factor
$F_\pi (q^2)$ is taken from the $e^+ e^- \to \pi^+ \pi^-$ cross-section
(the sign of $F_\pi(q^2)$ is not determined).

\begin{figure}[t]
\vspace{.4cm}
\includegraphics[width=2.8in]{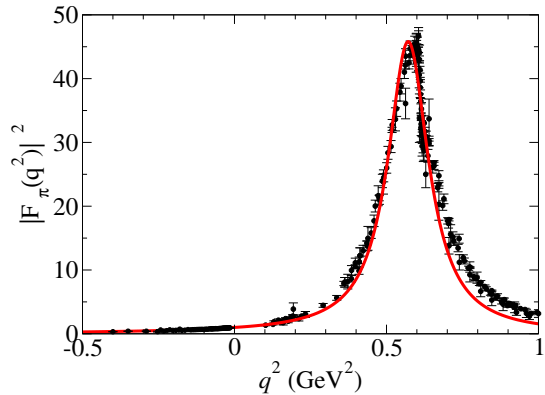}
\caption{\footnotesize
Fit to $|F_\pi (q^2)|^2$ data using
Eq.~(\ref{eqFpiB}).
The data are from Refs.~\cite{PionFF_data_TL,PionFF_data_SL}.
}
\label{figFpi2}
\end{figure}

The function $F_\pi (q^2)$ is well described by a simple monopole form
as $F_\pi(q^2)= \frac{\alpha}{\alpha -q^2 - i \beta}$, where $\alpha$ is
a cutoff squared and $\beta$ is proportional to a constant width.  An
alternative expression for $F_\pi (q^2)$, that replaces the Iachello
form $F_\rho$ is,
\begin{equation}
F_\pi (q^2)=   \frac{\alpha}{\alpha - q^2
- \frac{1}{\pi} \beta q^2 \log \frac{q^2}{m_\pi^2}
+ i \beta q^2}.
\label{eqFpiB}
\end{equation}
Eq.~(\ref{eqFpiB}) simulates the effect of the $\rho$ pole with an
effective width regulated by the parameter $\beta$.  Note that also
Eq.~(\ref{eqFpiB}) has a form similar to the function $F_\rho$ of the
Iachello model given by Eq.~(\ref{eqFrho}). In particular, when
$\alpha \to m_\rho^2$ and $\beta \to \frac{\Gamma_\rho^0}{m_\pi}$, we
recover Eq.~(\ref{eqFrho}).  The advantage of Eq.~(\ref{eqFpiB}) over
Eq.~(\ref{eqFrho}) is that $\alpha$ and $\beta$ can be adjusted
independently to the $|F_\pi|^2$ data.  The result for those parameters
from the fit in both time- and spacelike regions gives
\begin{equation}
\alpha = 0.696\;  \mbox{GeV}^2,  \hspace{.7cm} \beta= 0.178.
\label{eqParam}
\end{equation}
In the Iachello model (\ref{eqFrho}) one has $\beta \simeq 1.1$, a very
different value.  The fit is illustrated in Fig~\ref{figFpi2}.
The best fit selects $\alpha \simeq 0.7 \,\GeV^2$, which is larger than
$m_\rho^2 \simeq 0.6 \, \GeV^2$.
However, in the best fit to the data, the value of $\alpha$ is
{\it corrected} by the logarithmic counterterm in the denominator of
Eq.~(\ref{eqFpiB}), that pushes the maximum of $|F_\pi(q^2)|^2$ to the
correct position, $q^2 \simeq 0.6$ GeV$^2$.  In the Iachello model,
since $\beta \simeq 1.1$, the correction is too strong, and the maximum
moves to $q^2 \simeq 0.3$ GeV$^2$, differing significantly from the
$|F_\pi (q^2)|^2$ data.

To describe the physics associated with the $\rho$-meson, we restricted
the fit to $q^2 < 0.6 \, \GeV^2$, which causes a less perfect
description of $F_{\pi}$ at the right side of the peak.  However
increasing $q^2$ beyond that point slightly worsens the fit. This
probably indicates that although the $\omega$ width is small, there may
be some interference from the ${\omega}$ mass pole, and that the
parameters $\alpha$ and $\beta$ account for these interference effects.
Although the spacelike data was also included in the fit, the final
result is insensitive to the spacelike constraints.  We obtain also a
good description of the spacelike region (examine the region $q^2< 0$
GeV$^2$ in Fig~\ref{figFpi2}).  The full extension of the region where a
good description is achieved is $-1 \, \GeV^2 < q^2 < 1\, \GeV^2$.

A similar quality of the fit is obtained with both a constant width or a
$q^2$-dependent $\rho$-width.  However a better fit can be obtained with
a more complex $q^2$-dependence, which accounts better for the
$\omega$-meson pole effect, as shown in previous
works~\cite{Connell97,Donges94}.  Since this work is meant to probe the
quality of the results that one can obtain for the transitions form
factors, the simple analytic form of Eq.~(\ref{eqFpiB}) suffices for
$F_\pi(q^2)$.

\begin{figure}[t]
\vspace{.6cm}
\includegraphics[width=2.8in]{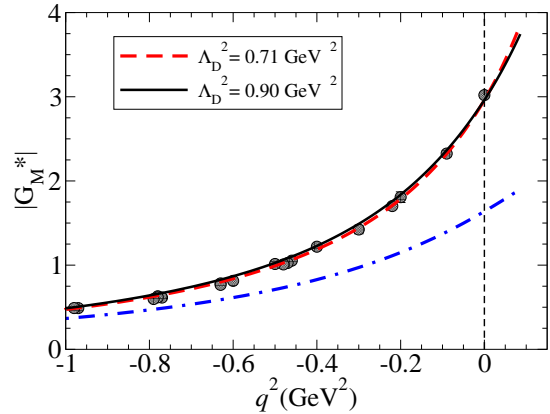}
\caption{Results for $|G_M^\ast(q^2)|$ for the covariant spectator quark
  model combined with the pion cloud contribution from
  Eq.~(\ref{eqGMmc}).  The data are from Refs.~\cite{GM-Data}.  The
  dashed-dotted-line is the contribution from the core~\cite{Timelike}.
}
\label{figGMTL}
\end{figure}

In addition, the covariant spectator quark model built from this
function describes well the $\Delta \to \gamma^\ast N$ form factor in
the spacelike region as shown in Fig.~\ref{figGMTL}.  Using the best fit
of $F_\pi$ given by the parameters (\ref{eqParam})
we can calculate the pion cloud contribution $G_M^\pi(q^2)$ through
Eq.~(\ref{eqGMmc}), and consequently the result for
$G_M^\ast(q^2,M_\Delta)$.  For the parameters $\lambda_\pi$ and
$\Lambda_\pi^2$ we use the results of the previous works
$\lambda_\pi= 0.441$ and $\Lambda_\pi^2 = 1.53 \, \GeV^2$, obtained from
the comparison of the constituent quark model to the lattice QCD data
and experimental data~\cite{NDeltaD,LatticeD,Timelike}.

In Fig.~\ref{figGMTL} we present the result of our model for
$|G_M^\ast(q^2,W)|$ for the case $W=M_\Delta$.  In that case the
imaginary contribution (when $q^2 > 0$) is very small and the results
can be compared with the spacelike data ($q^2 < 0$).  In the figure the
dashed-dotted-line indicate the result for $G_M^B(q^2,M_\Delta)$
discussed in a previous work~\cite{Timelike}.

In the same figure we show the sensitivity to the cutoff $\Lambda_D$ of
the pion cloud model, by taking the cases $\Lambda_D^2 = 0.71 \, \GeV^2$
and $\Lambda_D^2= 0.90 \, \GeV^2$.  They are are consistent with the
data, although the model with $\Lambda_D^2 = 0.71 \,\GeV^2$ gives a
slightly better description of the data. The two models are also
numerically very similar to the results of Ref.~\cite{Timelike} for
$W= M_\Delta$. For higher values of $W$ the results of the present
model and the ones from Ref.~\cite{Timelike} will differ.

Although the model with $\Lambda_D^2 = 0.71 \,\GeV^2$ gives a (slightly)
better description of the spacelike data, for the generalization to the
timelike region it is better to have a model with large effective
cutoffs when compared with the scale of the $\rho$ meson pole (the
$\rho$ mass $m_\rho$). This is important to separate the effects of the
physical scales from the effective scales (adjusted cutoffs).

\section{Results}

\begin{figure}[t]
\vspace{.4cm}
\includegraphics[width=2.8in]{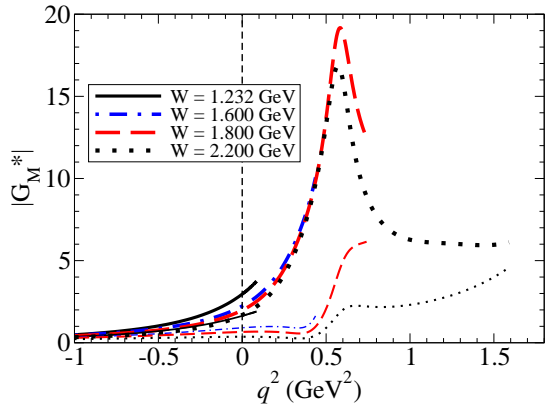}
\caption{ Results for $|G_M^\ast (Q^2)|$ for $W=1.232 \,\GeV$,
  $W=1.6 \,\GeV$, $W=1.8 \,\GeV$ and $W=2.2 \,\GeV$.  The thick lines
  indicate the final result.  The thin lines indicate the contribution
  of the core.}
\label{figGM-W}
\end{figure}

The results for $|G_M^\ast (q^2)|$ from the covariant spectator quark
model for the cases $W=1.232 \,\GeV$, $W=1.6 \,\GeV$, $W=1.8 \, \GeV$,
and $W=2.2 \,\GeV$ are presented in Fig.~\ref{figGM-W}. The thin lines
represent the contribution from the bare quark core component of the
model, and the thick line the sum of bare quark and pion cloud
contributions.

In the figure  the results for each value $W$
are restricted by the timelike kinematics through the condition
$q^2  \le (W - M)^2$,
since the nucleon and the resonance (with mass $W$) are treated
both as being on their mass shells.
Therefore the form factor covers an increasingly larger region on the
$q^2$ axis, as $W$ increases.
See Ref.~\cite{Timelike} for a complete discussion.

The figure illustrates well the interplay between the pion cloud and the
bare quark core components. The pion cloud component is dominating in
the region near the $\rho$ peak. Away from that peak it is the bare
quark contribution that dominates. The flatness of the $W=2.2 \, \GeV$
curve for $q^2 > 1 \,\GeV^2$ is the net result of the falloff of the
pion cloud and the rise of the quark core terms. In addition, the figure
shows that dependence on $W$ yields different magnitudes at the peak,
and we recall that this dependence originates from the bare quark core
contribution alone.
This bare quark core contribution is mainly
the consequence of the VMD parameterization
of the quark current where there is an interplay between
the effect of the  $\rho$ pole and a term that behaves as a constant
for intermediate values of $q^2$ (see Appendix~\ref{appQuarkFF}).

We will discuss now the results for the widths
$\Gamma_{\gamma^\ast N}(q,W)$ of the $\Delta$ Dalitz decay, and for the
$\Delta$ mass distribution $g_\Delta (W)$.

\subsection{$\Delta$ Dalitz decay}

\begin{figure}[t]
\vspace{.4cm}
\includegraphics[width=2.8in]{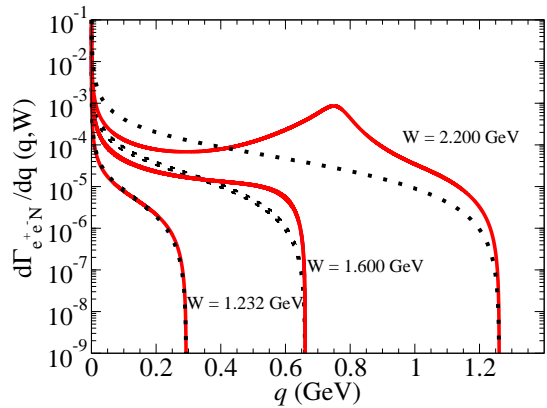}
\caption{Results for $\frac{d \Gamma_{e^+ e^- N}}{d q} (q,W)$ for three
  different values of energies $W$.  The solid line is the result of our
  model.  The dotted line is the result of the constant form factor
  model.}
\label{figDGamma}
\end{figure}

The width associated with the $\Delta$ decay into $\gamma^\ast N$ can be
determined from the $\Delta \to \gamma^\ast N$ form factors for the
$\Delta$ mass $W$.
Assuming the dominance of the magnetic dipole
form factors over the other two transition form factors,
we can write~\cite{Timelike,Frohlich10,Krivoruchenko01}
\begin{equation}
\begin{split}
\Gamma_{\gamma^\ast N} (q,W)= &
\frac{\alpha}{16} \frac{(W+M)^2}{M^2 W^3}  \\
& \times\sqrt{y_+ y_-} y_- |G_M^\ast (q^2,W)|,
\label{eqGamma_gX}
\end{split}
\end{equation}
where $q=\sqrt{q^2}$, $\alpha \simeq 1/137$ is the fine-structure
constant and $y_\pm= (W\pm M)^2 - q^2$.

At the photon point ($q^2=0$), in particular, we obtain the
$\Gamma_{\gamma N}$ in the limit $q^2=0$ from
Eq.~(\ref{eqGamma_gX})~\cite{Frohlich10,Wolf90,Krivoruchenko02}
\begin{equation}
\Gamma_{\gamma N} (W)=\Gamma_{\gamma^\ast N} (0,W).
\label{eq-gN}
\end{equation}

We can also calculate the derivative of the Dalitz decay width
$\Gamma_{e^+ e^- N} (q,W)$ from the function
$\Gamma_{\gamma^\ast N} (q,W)$ using the
relation~\cite{Frohlich10,Wolf90,Krivoruchenko02,Krivoruchenko01}
\begin{equation}
\begin{split}
\Gamma^\prime_{e^+ e^- N}(q,W) & \equiv
\frac{\dd \Gamma_{e^+ e^- N}}{\dd q} (q,W) \\
&= \frac{2 \alpha}{3 \pi q} \Gamma_{\gamma^\ast N} (q,W).
\end{split}
\end{equation}
The Dalitz decay width $\Gamma_{e^+ e^- N} (q,W)$
is given by
\begin{equation}
\Gamma_{e^+ e^- N} (W) =
\int_{2 m_e}^{W-M} \Gamma^\prime_{e^+ e^- N}(q,W) \, \dd q,
\label{eq-geeN}
\end{equation}
where $m_e$ is the electron mass.
Note that the integration holds for the interval
$4m_e^2 \le q^2 \le (W-M)^2$, where the lower limit is the minimum value
necessary to produce an $\mathrm{e}^+ \mathrm{e}^-$ pair, and $(W-M)^2$
is the maximum value available in the $\Delta \to \gamma^\ast N$ decay
for a given $W$ value.

The results for $\frac{\dd \Gamma_{e^+ e^- N}}{\dd q} (q,W)$ for several
mass values $W$ (1.232, 1.6 and 2.2 GeV) are presented in
Fig.~\ref{figDGamma}.  These results are also compared to the
calculation given by the constant form factor model, from which they
deviate considerably.

\begin{figure*}[t]
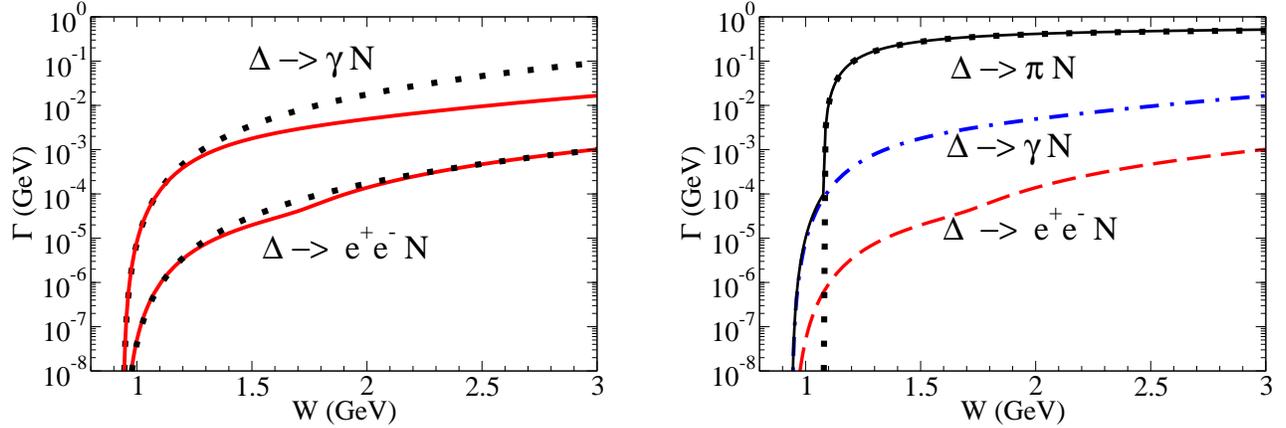

\vspace{.4cm}
\includegraphics[width=3.1in]{Gamma012_mod4b}
\hspace{.8cm}
\includegraphics[width=3.1in]{GammaTotal4-Manley}
\caption{\footnotesize{
Results for the partial widths as function of $W$.
At left: partial widths (solid line) for $\Delta \to \gamma N$
and   $\Delta \to e^+ e^- N$, compared with
the constant form factor model (dotted line).
At right: the partial widths are compared
with the $\Delta \to \pi N $ width (dotted line) and
with the sum of all widths (thin solid line).
}}
\label{figGamma}
\end{figure*}

\begin{figure*}[t]
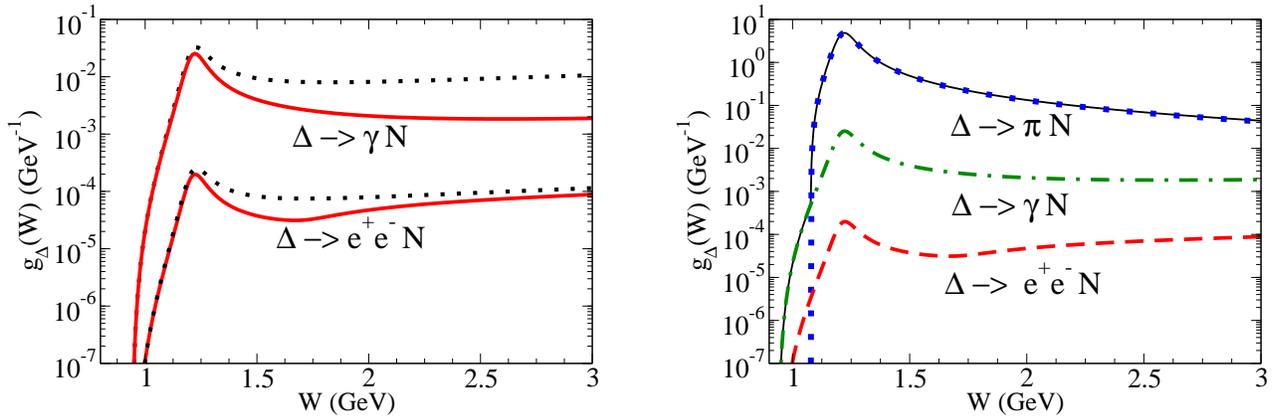

\vspace{.4cm}
\includegraphics[width=3.1in]{gtilT4}
\hspace{.8cm}
\includegraphics[width=3.1in]{gtil4}
\caption{ Results for $g_\Delta(W)$ and the partial contributions
  $g_{\Delta \to \pi N}(W)$, $g_{\Delta \to \gamma N}(W)$ and
  $g_{\Delta \to e^+ e^- N}(W)$.  At left: $g_{\Delta \to \gamma N}(W)$
  and $g_{\Delta \to e^+ e^- N}(W)$ in comparison with constant form
  factor model (dotted line).  At right: all contributions compared with
  the total $g_\Delta (W)$ (thin solid line).}
\label{figGtil}
\end{figure*}

Also, the $\Delta$ decay width can be decomposed at tree level into
three independent channels
\begin{equation}
\Gamma_{\rm tot}(W)=
\Gamma_{\pi N} (W) + \Gamma_{\gamma N} (W) +
\Gamma_{e^+ e^- N} (W),
\end{equation}
given by
the decays $\Delta \to \pi N$, $\Delta \to \gamma N$
and $\Delta \to e^+ e^- N$.
The two last terms are described respectively
by Eqs.~(\ref{eq-gN}) and (\ref{eq-geeN}).
The $\Gamma_{\pi N}$ term can be
parameterized as in~\cite{Manley92,Buss11}
\begin{equation}
\Gamma_{\pi N}(W)=
\frac{M_\Delta}{W}
\left( \frac{q_\pi (W)}{q_\pi(M_\Delta)} \right)^3
\frac{\kappa^2 + q_\pi^2 (M_\Delta)}{\kappa^2 + q_\pi^2 (W)}
\Gamma_{\pi N}^0,
\label{eqGMtot}
\end{equation}
where $\Gamma_{\pi N}^0$ is the $\Delta \to \pi N$ partial width for the
physical $\Delta$, $q_\pi (W)$ is the pion momentum for a $\Delta$ decay
with mass $W$, and $\kappa$ a cutoff parameter.
Following Refs.~\cite{Weil12,Weil14} we took $\kappa =0.197$ GeV. The
present parameterization differs from other forms used in the
literature~\cite{Frohlich10,Wolf90} and from our previous
work~\cite{Timelike}.

The results for the partial widths as functions of the mass $W$ are
presented in Fig.~\ref{figGamma}.  On the left panel we compare
$\Gamma_{\gamma N}$ and $\Gamma_{e^+ e^- N}$ with the result of the
constant form factor model.  On the right panel we present the total
width $\Gamma_{\rm tot} (W)$ as the sum of the three partial widths.

\subsection{$\Delta$ mass distribution}

To study the impact of the $\Delta$ resonance propagation in nuclear
reactions like the $NN$ reaction, it is necessary to know the $\Delta$
mass distribution function $g_\Delta(W)$.  As discussed before, $W$ is
an arbitrary resonance mass that may differ from the resonance pole mass
($M_\Delta$).  The usual ansatz for $g_\Delta$ is the relativistic
Breit-Wigner distribution-\cite{Timelike,Frohlich10}
\begin{equation}
g_\Delta (W)= A \frac{W^2 \Gamma_{\rm tot}(W)}{(W^2 - M_\Delta^2)^2 +
W^2 \left[\Gamma_{\rm tot} (W)  \right]^2},
\label{eqSF}
\end{equation}
where $A$ is a normalization constant determined by
$\int g_\Delta(W) \dd W =1$ and the total width $\Gamma_{\rm tot}(W)$
(\ref{eqGMtot}).

The results for $g_\Delta(W)$ and the partial contributions
\begin{figure*}[t]
\includegraphics[width=\textwidth]{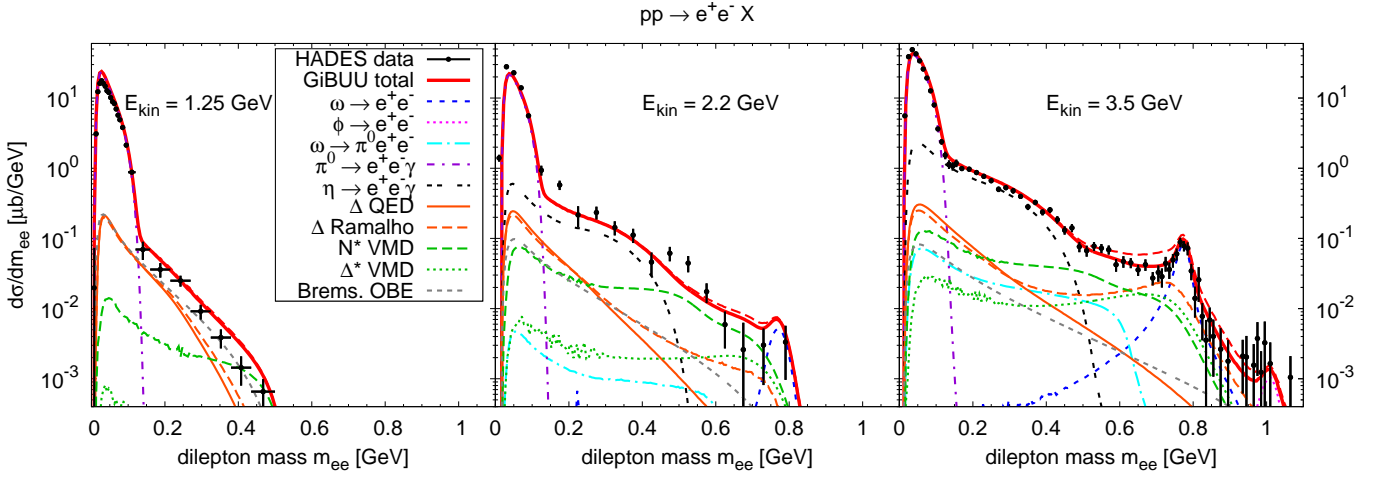}
\caption{Transport-model calculations of dilepton mass spectra $\dd \sigma/\dd m_{ee}$ from
proton-proton collisions $(pp \to e^+ e^- X)$ at three different beam energies,
with and without
a $\Delta \to \gamma^\ast N$ form factor,
compared to experimental data measured with the HADES detector
\cite{Agakishiev09,Agakishiev12,HADES12}.}
\label{figSigma}
\end{figure*}
\begin{alignat}{2}
& g_{\Delta \to \gamma N}(W)= \frac{\Gamma_{\gamma N}(W)}{\Gamma_{\rm tot}(W)}
g_\Delta(W), \\
&
\label{eqSFdil}
g_{\Delta \to e^+ e^- N}(W)= \frac{\Gamma_{e^+ e^-  N}(W)}{\Gamma_{\rm tot}(W)}
g_\Delta(W), \\
&
g_{\Delta \to \pi N}(W)= \frac{\Gamma_{\pi N}(W)}{\Gamma_{\rm tot}(W)}
g_\Delta(W),
\end{alignat}
are presented in Fig.~\ref{figGtil}.  The results are also compared with
the constant form factor model.

\subsection{Dilepton production from $NN$ collisions}

The $\Delta \to \gamma^\ast N$ magnetic dipole form factor in the
timelike region is known to have a significant influence on dilepton
spectra. Therefore we show in Fig.~\ref{figSigma} a transport-model
calculation of the inclusive dielectron production cross section
$\dd \sigma/\dd m_{\text{ee}}$ for proton-proton collisions
$(\text{pp} \to \mathrm{e}^+ \mathrm{e}^- X)$, where $m_{\text{ee}} = q$. These
results have been obtained with the GiBUU model \cite{Buss11,Weil12} for
three different proton beam energies and are compared to experimental
data measured with the HADES detector
\cite{Agakishiev09,Agakishiev12,HADES12}.
Except for the contribution of the $\Delta$ Dalitz decay, the
calculations are identical to those presented in an earlier publication
\cite{Weil14}. The $\Delta$ Dalitz decay is shown in two variants, once
with a constant form factor fixed at the photon point (i.e., in 'QED'
approximation) and once using the form-factor model described in the
preceding sections.

\begin{figure}[b]
\vspace{1cm}
\includegraphics[width=3.1in]{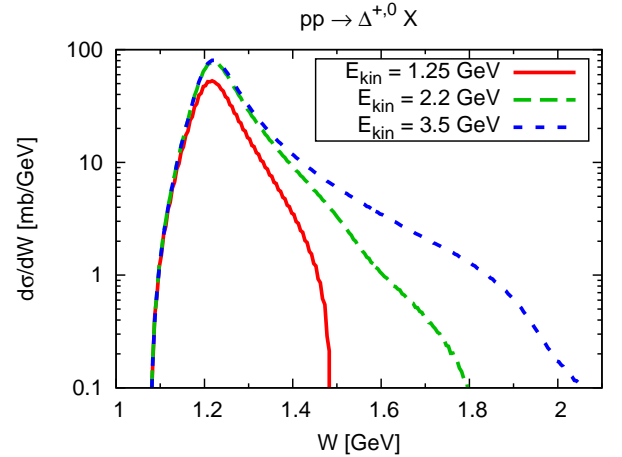}
\caption{Mass distribution of produced $\Delta^{+,0}$ baryons in GiBUU
  simulations, for pp collisions at three different beam energies.}
\label{figWdistr}
\end{figure}

At the lowest beam energy of 1.25 GeV, the produced $\Delta$ baryons are
close to the pole mass and therefore the results with and without the
form factor are very similar. At higher beam energies, however, the
model for the $\Delta \to \gamma^\ast N$ form factor has a much larger
impact, because higher values of $W$ are reached, where the form factor
deviates strongly from the photon point value. In Fig.~\ref{figWdistr}
we illustrate the influence of $W$ by showing the $W$ distribution of
produced $\Delta^{+,0}$ baryons in the GiBUU simulations. We note
that several different processes contribute to the inclusive
  $\Delta^{+,0}$ production, such as
$NN\rightarrow N\Delta,\; \Delta\Delta,\; \Delta N^*$ etc., each of
which will produce a different $W$ distribution due to different
kinematics and phase space.
Furthermore it should be remarked that
  the tails of this distribution, just as the $\Delta$ spectral function
  in Eq.~(\ref{eqSF}), depend significantly on the specific
  parameterization of the hadronic width for $\Delta\rightarrow\pi N$.
  However, for electromagnetic observables as shown in
  Fig.~\ref{figSigma}, the dependence on the hadronic width is very
  weak, since in Eq.~(\ref{eqSFdil}) the total width cancels out in the
  numerator and only stays in the denominator.

\begin{figure*}[htb]
\includegraphics[width=\textwidth]{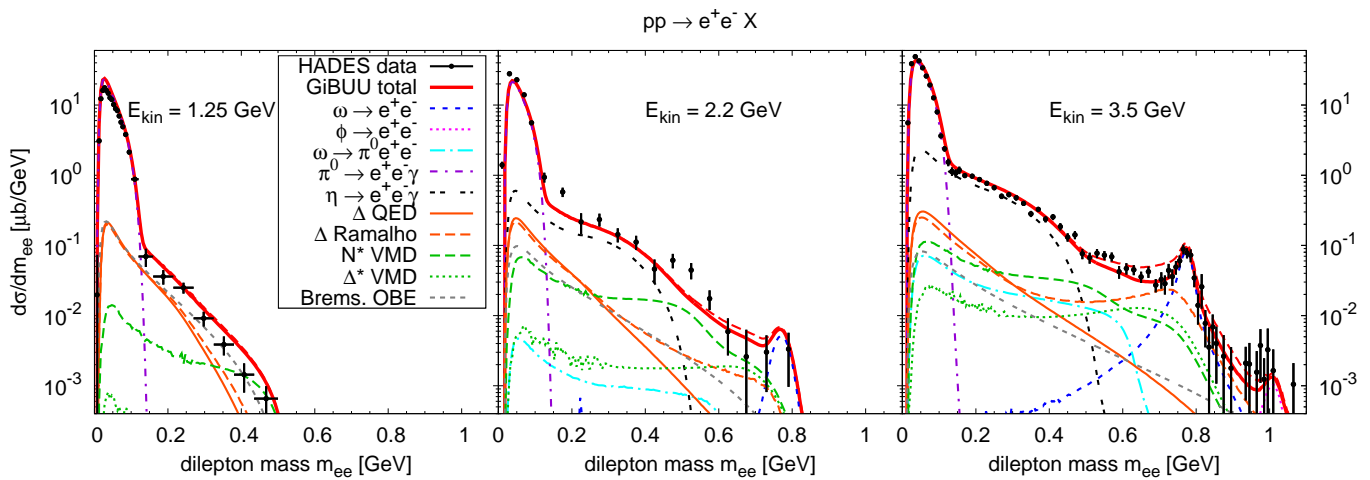}
\caption{Modified calculations of dilepton mass spectra
  $\dd \sigma/\dd m_{\text{ee}}$ from proton-proton collisions
  $(\text{pp} \to \text{e}^+ \text{e}^- X)$, using reduced $R\rightarrow\rho N$ branching
  ratios for two resonances (see text).}
\label{figSigma_mod}
\end{figure*}

Coming back to Fig.~\ref{figSigma}, it should be noted that the choice
of the form factor has little influence on the overall agreement of the
total dilepton spectrum with the experimental data at the two lowest
beam energies, because the influence of the form factor is weak or the
$\Delta$ contribution is small compared to other channels.  At the
highest beam energy of 3.5 GeV, however, the choice of the form factor
does have an impact on the total spectrum for masses above 600 MeV.
While the constant-form-factor result combined with the other channels
from GiBUU shows a good agreement with the data, using the $q^2$
dependent form factor results in a slight overestimation of the data,
which is most severe for masses of around 700 MeV. However, we note that
the $\Delta$ contribution by itself does not overshoot the data. Only in
combination with the other channels (in particular the heavier baryons,
such as $N^*$ and $\Delta^*$) the overestimation is seen.

There could be various reasons for this enhancement over the data, but
we want to mention here only the two most likely ones. One could lie in
the form factor itself, more precisely in the omission of an $W$
dependence of the overall weight $\lambda_{\pi}$ for the pion
cloud. This parameter for the weight of the pion cloud should probably
depend on $W$. If the two diagrams (a) and (b) of the pion cloud
contribution would decrease simultaneously with $W$, as we can expect from
the drop of the $m_\pi/W$ ratio, this could potentially cure the
observed overestimation.

On the other hand, the reason for the disagreement could also be found
in the other channels that are part of the transport calculation. In
particular the contributions of the higher baryonic resonances ($N^*$
and $\Delta^*$) are subject to some uncertainties. These resonance
contributions were recently investigated via exclusive pion production
at 3.5 GeV with the HADES detector \cite{Agakishiev14}, which showed
that the GiBUU model does a rather good job in describing the resonance
cocktail for the exclusive channels (with some minor deviations).
However, there are also significant non-exclusive channels for pion and
dilepton production at this energy. Moreover, the form factors of the
higher resonances are a matter of debate (they are treated in a
strict-VMD assumption in the calculation).

  It was remarked in \cite{Agakishiev14} that some of the branching
  ratios for $R\rightarrow\rho N$, which directly influence the dilepton
  yield via the VMD assumption, might be overestimated in GiBUU, in
  particular for the $N^*(1720)$ and the $\Delta^*(1905)$. Both have a
  very large $\rho N$ branching ratio of 87\% in GiBUU \cite{Weil12} (as
  adopted from \cite{Manley92}) and also in the current PDG database
  these branching ratios are listed with rather large values
  \cite{PDG14}, which are essentially compatible with the GiBUU values.
  However, some recent partial-wave analyses \cite{BonnGatchina,KSU}
  claim much smaller values for these branching ratios, showing some
  tension with the PDG and GiBUU values. We show in
  Fig.~\ref{figSigma_mod} the effect of using smaller values for these
  branching ratios on the dilepton spectra, adopting the upper limits
  from the Bonn-Gatchina analysis \cite{BonnGatchina} (as given in
  \cite{Agakishiev14}), namely 10\% for $N^*(1720)\rightarrow\rho N$ and
  42\% for $\Delta^*(1905)\rightarrow\rho N$. We note that the values in
  \cite{KSU} are even smaller. As seen in Fig.~\ref{figSigma_mod}, this
  change indeed reduces the contributions from the $N^*$ and $\Delta^*$
  resonances by a fair amount, in particular in the high-mass region
  ($m_{ee}>600$ MeV). This improves the agreement with the highest data
  points at 2.2 GeV, and it also mitigates the overshooting over the data
  at 3.5 GeV when the
  $\Delta \to \gamma^\ast N$ form factor is used, but it does not
  fully cure it.

  Thus it is quite likely that the remaining excess is caused by the
  negligence of the $W$ dependence in the pion cloud contribution of the
  form factor. A more detailed investigation of the $W$ dependence of
  the pion cloud is planned in a further study that will analyze all
  these aspects.

\section{Summary and conclusions}

In this work we present a new covariant model for the
$\Delta \to \gamma^\ast N$ transition in the timelike region.
The model is based on the combination of valence quark and meson cloud
degrees of freedom.  The bare quark contribution was calibrated
previously to lattice QCD data.
One of the pion cloud components is fitted to the pion electromagnetic
form factor $F_\pi$ (with the fit being almost insensitive to the
spacelike data and strongly dependent on the timelike data) and the
other, associated with intermediate octet/decuplet baryon states,
parameterized by an effective cutoff $\Lambda_D$.

Our model induces a strong effect on the
$\Delta \to \gamma^\ast N$ magnetic dipole form factor in the region
around the $\rho-$meson pole (where the magnitude is about four times
larger than at $q^2=0$). This effect was missing in the frequently used
Iachello model. The pion cloud effects dominate in the region
$q^2 \le 1.5 \, \GeV^2$. For larger $q^2$ the effects of the valence
quark became dominant, and the $q^2$-dependence is smoother.  At low
energies, the new form factor has little influence on the overall
agreement of the total dilepton spectrum in $NN$ collisions with the
experimental data, and no large difference between our new model and the
VMD model is seen. However at the highest beam energy of 3.5 GeV, the
choice of the form factor does affect the total spectrum for masses
above 600 MeV.

Measurements of independent channels, for instance exclusive pion
induced $\Delta$ production data, can help to better constrain the pion
cloud contribution. The methods presented in this work can in principle
be extended to higher mass resonances as $N^\ast(1440)$, $N^\ast(1520)$,
$N^\ast(1535)$, $N^\ast(1710)$ and
$\Delta^\ast(1600)$, for which there are already
predictions of the covariant spectator quark
model~\cite{N1520,Resonances} in the spacelike region. The calculation
of the $N^\ast(1520)$ form factors in the timelike region~\cite{N1520TL},
extending the results from Ref.~\cite{N1520} is already under way.


\begin{acknowledgments}
The authors thank Marcin Stolarski and Elmar
Biernat for the information about the pion electromagnetic form factors.
G.R.~was supported by the Brazilian Ministry of Science, Technology and
Innovation (MCTI-Brazil).
M.T.P.~received financial support from Funda\c{c}\~ao
para a Ci\^encia e a Tecnologia (FCT) under Grants
Nos.~PTDC/FIS/113940/2009, CFTP-FCT
(PEst-OE/FIS/U/0777/2013) and POCTI/ISFL/2/275.
This work was also partially supported by the European Union
under the HadronPhysics3 Grant No.~283286.
J.W.~acknowledges funding of a Helmholtz Young Investigator Group
VH-NG-822 from the Helmholtz Association and GSI.
\end{acknowledgments}

\appendix

\section{Quark form factors}
\label{appQuarkFF}

We use a parameterization of the quark isovector form factors motivated
by VMD~\cite{Nucleon,Lattice,LatticeD}
\begin{equation}
\begin{split}
& f_{1-} (q^2)= \lambda_q
+ (1-\lambda_q)
\frac{m_\rho^2}{m_\rho^2- q^2}
- c_- \frac{M_h^2 q^2}{(M_h^2-q^2)^2} \\
&f_{2-} (q^2)= \kappa_-
\left\{
d_-
\frac{m_{\rho}^2}{m_{\rho}^2- q^2} +
(1- d_-) \frac{M_h^2}{M_h^2-q^2}\right\},
\label{eqQFF}
\end{split}
\end{equation}
where $m_{\rho}= 775$ MeV is the $\rho$-meson mass,
$M_h$ is the mass of an effective heavy vector meson, $\kappa_-$ is the quark
isovector anomalous magnetic moment, $c_-$, $d_-$ are mixture
coefficients, and $\lambda_q$ is a parameter related with the quark
density number in the deep inelastic limit~\cite{Nucleon}.
The term in $M_h$, where $M_h=2M$,
simulates the effects of the heavier mesons
(short range physics)~\cite{Nucleon},
and behaves as a constant for values of
$q^2$ much smaller than $4M^2$.
The width associated with the pole $q^2=M_h^2$ is discussed in the
Appendix~\ref{appRegularize}.

The $\rho$ pole appears when one assumes a stable $\rho$ with zero decay
width $\Gamma_\rho=0$.  For the extension of the quark form factors to
the timelike regime we consider therefore the replacement
\begin{equation}
\frac{m_\rho^2}{m_\rho^2 -q^2}
\to
\frac{m_\rho^2}{m_\rho^2 -q^2 - \ii m_\rho \Gamma_\rho(q^2)}.
\end{equation}
On the r.h.s.~we introduce $\Gamma_\rho$ the $\rho$ decay width as a
function of $q^2$.

The function $\Gamma_\rho(q^2)$ represents the $\rho \to 2 \pi$ decay
width for a virtual $\rho$ with momentum
$q^2$~\cite{Gounaris68,Connell97}
\begin{equation}
\Gamma_{\rho} (q^2)=
\Gamma_{\rho}^0 \frac{m_\rho^2}{q^2}
\left(\frac{ q^2 - 4 m_\pi^2}{m_\rho^2- 4 m_\pi^2} \right)^{\frac{3}{2}}
\theta(q^2 - 4 m_\pi^2),
\end{equation}
where $\Gamma_{\rho}^0 = 0.149$ GeV.

\section{Regularization of high momentum poles}
\label{appRegularize}

For a given $W$ the squared momentum $q^2$ is limited by the kinematic
condition $q^2 \le (W-M)^2$.  Then, if one has a singularity at
$q^2=\Lambda^2$, that singularity will appear for values of $W$ such
that $\Lambda^2 \le (W-M)^2$, or $W \ge M + \Lambda$.

To avoid a singularity at $q^2=\Lambda^2$, where $\Lambda^2$ is any of
the cutoffs introduced in our pion cloud parameterizations, and quark
current (pole $M_h$) we implemented a simple procedure. We start with
\begin{equation}
\frac{\Lambda^2}{\Lambda^2 - q^2}
\to \frac{\Lambda^2}{\Lambda^2 - q^2 - \ii \Lambda \Gamma_X(q^2)},
\end{equation}
where
\begin{equation}
\Gamma_X(q^2)= 4 \Gamma_X^0
\left( \frac{q^2}{q^2+ \Lambda^2}
\right)^2
\theta(q^2),
\label{eqGammaX}
\end{equation}
In the last equation $\Gamma_X^0$ is a constant
given by  $\Gamma_X^0 = 4 \Gamma_\rho^0\simeq 0.6$ GeV.

In Eq.~(\ref{eqGammaX}) the function $\Gamma_X(q^2)$ is defined such
that $\Gamma_X(q^2)=0$ when $q^2 < 0$.  Therefore the results in
the spacelike region are kept unchanged.
For $q^2= \Lambda^2$ we obtain $\Gamma_X = \Gamma_X^0$, and for very
large $q^2$ it follows $\Gamma_X \simeq 4 \Gamma_X^0$.
Finally the value of $\Gamma_X^0$ was chosen to avoid very narrow peaks
around $\Lambda^2$.

While the width $\Gamma_\rho(q^2)$ associated with the $\rho$-meson pole
in the quark current is nonzero only when $q^2 > 4 m_\pi^2$, one has for
$\Gamma_X(q^2)$ nonzero values also in the interval
$4 m_\pi^2 > q^2 >0$.  However, the function $\Gamma_X(q^2)$ changes
smoothly in that interval and its values are negligible.

This procedure was used in Ref.~\cite{Timelike,Weil12} for the
calculation of the $\Delta \to \gamma^\ast N$ form factors in the
timelike regime.  In the present case the emerging singularities for
$W > M + \Lambda_D \simeq 1.84 \,\GeV$ are avoided, and for
$W < 1.84 \,\GeV$, the results are almost identical to the ones without
regularization. The suggested procedure avoids the singularities at high
momentum and at the same time preserves the results for low momentum.
In the cases considered the high $q^2$ contributions are suppressed and
the details of regularization procedure are not important.


\end{document}